\documentclass[aps,prl,twocolumn,longbibliography,superscriptaddress]{revtex4-2}

\usepackage{amsfonts}
\usepackage{amsmath}
\usepackage{graphicx}
\usepackage{dsfont}
\usepackage{diagbox}
\usepackage{array}
\usepackage{amssymb}
\usepackage{bbm}
\usepackage{bm}
\usepackage{float}
\usepackage{subfigure}
\usepackage{MnSymbol}
\usepackage{url}
\usepackage{times}
\usepackage{manfnt}
\usepackage{booktabs}
\usepackage{xcolor}
\definecolor{darkblue}{HTML}{004D6B}
\definecolor{darkred}{HTML}{8c1515}
\definecolor{darkgreen}{HTML}{006400}
\usepackage{hyperref}
\hypersetup{
    pdftitle={measuringIsingState},
	colorlinks=true,
	urlcolor=darkred,
	citecolor=darkblue,
	linkcolor=darkred,
	breaklinks
}
\pagecolor{white}
\usepackage{physics}
\usepackage{enumitem}
\usepackage{wasysym}
\usepackage{tikz}
\makeatletter

\begin{document}

\title{Learning transitions in classical Ising models and deformed toric codes}

 \author{Malte Pütz}
 \affiliation{Institute for Theoretical Physics, University of Cologne, Z\"ulpicher Straße 77, 50937 Cologne, Germany}

\author{Samuel J. Garratt}
\affiliation{Department of Physics, University of California, Berkeley, California 94720, USA}

\author{Hidetoshi Nishimori}
\affiliation{Institute of Integrated Research, Institute of Science Tokyo, Nagatsuta-cho, Midori-ku, Yokohama 226-8503, Japan}

 \author{Simon Trebst}
 \affiliation{Institute for Theoretical Physics, University of Cologne, Z\"ulpicher Straße 77, 50937 Cologne, Germany}

 \author{Guo-Yi Zhu}
\email{guoyizhu@hkust-gz.edu.cn}
\affiliation{The Hong Kong University of Science and Technology (Guangzhou), Nansha, Guangzhou, 511400, Guangdong, China}

\date{\today}

%%%%%%%%%%%%%%%%%%%%%%%%%%%%%%%%%%%%%%%%%%%%%%%%%%%%%%%%%%%%%%%%%%%
\begin{abstract}
Conditional probability distributions describe the effect of learning an initially unknown classical state through Bayesian inference.
Here we demonstrate the existence of a \textit{learning transition}, having signatures in the long distance behavior of conditional correlation functions, in the two-dimensional classical Ising model. This transition, which arises when learning local energy densities, extends all the way from the infinite-temperature paramagnetic state down to the thermal critical state. The intersection of the line of learning transitions and the thermal Ising transition is a new tricritical point. Our model for learning also exactly describes the effects of weak measurements on ground states of frustration-free quantum Hamiltonians, which interpolate between the toric code and a paramagnet.
Notably, the location of the above tricritical point implies that the quantum memory defined by the degenerate ground states in the topological phase is robust to weak measurement, even when the initial state is arbitrarily close to the quantum phase transition separating topological and trivial phases.  
Our analysis uses a replica field theory combined with the renormalization group, and we chart out the 
phase diagram using a combination of tensor network and Monte Carlo techniques. 
Our methods can be extended to study the more general effects of learning on both classical and quantum states. The learning induced critical states can be realized in classical or quantum devices. 
\end{abstract}
%%%%%%%%%%%%%%%%%%%%%%%%%%%%%%%%%%%%%%%%%%%%%%%%%%%%%%%%%%%%%%%%%%%

\maketitle

%%%%%%%%%%%%%%%%%%%%%%%%%%%%%%%%%%%%%%%%%%%%%%%%%%%%%%%%%%%%%%%%%%%

%%%%%%%%%%%%%%%%%%%%%%%%%%%%%%%%%%%%%%%%%%%%%%%%%%%%%%%%%%%%%%%%%%%
\begin{figure*}[th!]
   \centering
   \includegraphics[width=\linewidth]{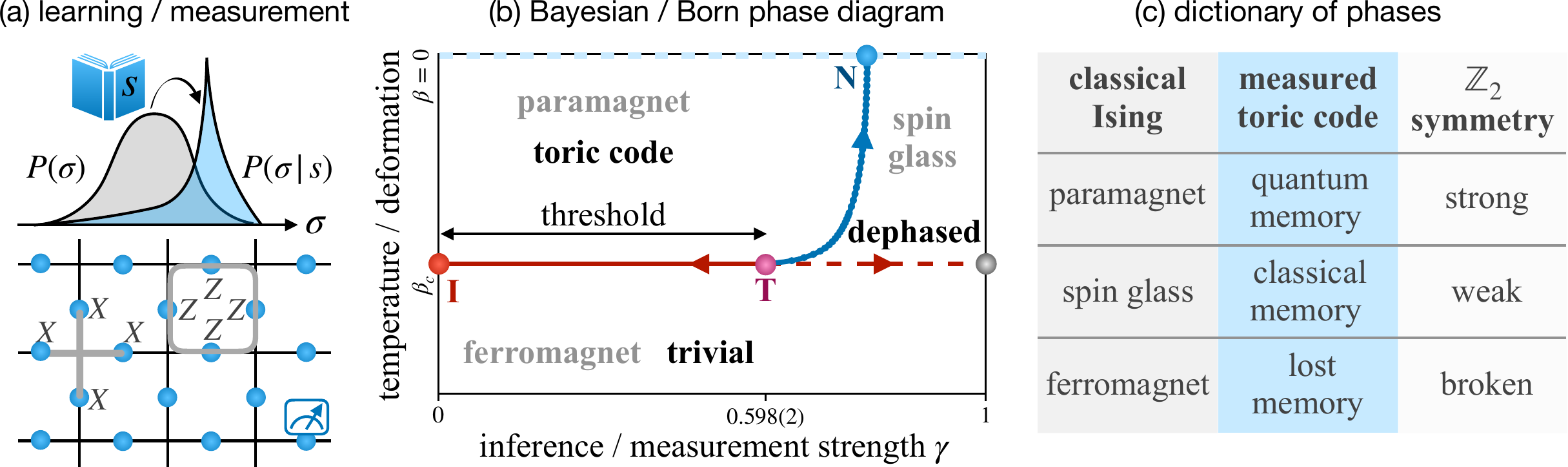} 
   \caption{ {\bf Schematic setup, numerical phase diagram, and  duality}.
   (a) For a classical Ising model with spins $\sigma_i$, learning $s_{ij}$ that is correlated with $\sigma_i \sigma_j$ (via a `strength’ $\gamma$) changes our state of knowledge from the Boltzmann distribution $P(\bm{\sigma}) \sim e^{-\beta E(\sigma)}$ to the conditional distribution $P(\bm{\sigma}|\bm{s})$. Similarly, when measuring a quantum system initially in a Rokhsar-Kivelson state $\ket{\psi}=\sum_{\bm{\sigma}}\sqrt{P(\bm{\sigma})}\ket{\bm{\sigma}}$, the post-measurement state takes the form $\ket{\psi(\bm{s})}=\sum_{\bm{\sigma}}\sqrt{P(\bm{\sigma}|\bm{s})}\ket{\bm{\sigma}}$. Here these states are Wegner-dual to those of deformed toric codes, where $\beta$ is the strength of the wavefunction deformation. 
   (b) Phase diagram according to the classical Bayesian rule or the quantum Born rule. The paramagnet (PM) and ‘spin glass’ (SG) phases correspond to short- and long-range nonlinear correlations $[\langle \sigma_i \sigma_j \rangle_{\bm{s}}^2] \equiv \sum_{\bm{s}} [\sum_{\bm{\sigma}} P(\bm{\sigma}|\bm{s}) \sigma_j \sigma_j]^2$, respectively, but both have short-range linear correlations $[\langle \sigma_i \sigma_j \rangle_{\bm{s}}]=\langle\sigma_i \sigma_j\rangle$, while the ferromagnet (FM) has long-range linear correlations. The three different universality classes of the transitions we study are indicated I (Ising), N (Nishimori) and T (tricritical). The red transition lines are straight at $\beta = \beta_c$, with Ising critical correlation $[\langle \sigma_i \sigma_j \rangle_{\bm{s}}]$. For $\gamma > \gamma_c(\beta)$ both SG and FM have long-range correlation $|\langle \sigma_i\sigma_j\rangle_{\bm{s}}|=O(1)$ and, as we cross the dashed line between them, at $\beta=\beta_c$ the distribution of $\bm{s}$ undergoes an Ising transition.
   (c) Relation between phases of the classical inference problem and the measured and deformed toric code, and symmetries of the Rokhsar-Kivelson classical-quantum state $\sum_{\bm{s}}P(\bm{s}) \ket{\bm{s}}\bra{\bm{s}} \otimes \ket{\psi(\bm{s})}\bra{\psi(\bm{s})}$ consisting of both the measurement record $\ket{\bm{s}}$ and post-measurement states $\ket{\psi(\bm{s})}$.}
    \label{fig:phasediagram}
\end{figure*}
%%%%%%%%%%%%%%%%%%%%%%%%%%%%%%%%%%%%%%%%%%%%%%%%%%%%%%%%%%%%%%%%%%%

In classical statistical mechanics the state of a many-body system is a high-dimensional probability distribution, and learning can be captured through Bayesian inference~\cite{Jaynes1957,Jaynes1957b,nishimori2001statistical,Zdeborova16inferencethreshold,Golan22info}: an initial probability distribution (e.g.\ a Boltzmann distribution) is updated to a sharper conditional distribution reflecting improved knowledge. This setting is dual
to one arising in many-body quantum mechanics: the measurement of an observable causes a back-action on a quantum many-body state described by Born's rule. There, a fundamental problem is to determine when it is that quantum information is robust to measurement and decoherence~\cite{Nielsen2010,Fisher2022reviewMIPT, Potter21review}. However, in both classical and quantum settings the change in the many-body state depends sensitively on the initial correlations. This sensitivity raises the question of when it is that the effects of learning (or measurements) are universal, depending only on qualitative features of the initial states and learning protocols.

Here, we first study the effects of learning on high-entropy probability distributions describing the states of correlated classical binary degrees of freedom, or `spins'. Our focus is on the two-dimensional classical Ising model on a square lattice, with our state of knowledge initially described by classical Gibbs states parametrized by inverse temperature $\beta$. We show that, throughout the entire high temperature phase $\beta < \beta_c$, and also at the thermal phase transition $\beta=\beta_c$, learning two-point correlations between neighboring spins leads to a sharp transition in our knowledge of long-distance correlations. These transitions occur only when the local entropy reduction, parameterized by a variable $\gamma$, exceeds a system-dependent $\gamma_c(\beta)$.

Such a transition has previously been identified \cite{Iba1999} only at infinite temperature $\beta=0$, and the fact that the learning transition with finite $\gamma$ extends down to the thermal phase transition implies a {\it novel tricritical point} in the $\beta-\gamma$ plane, at $\beta=\beta_c$ and $\gamma_T\equiv\gamma_c(\beta_c) > 0$. By adapting a replica field theory to describe the effects of learning, and by leveraging the renormalization group to describe this theory, we provide evidence that the above behavior is universal to two-dimensional systems with global $\mathbbm{Z}_2$ symmetry and when learning the values of local $\mathbbm{Z}_2$ symmetric observables. We then probe the learning transitions in the Ising model numerically, and we reveal that the critical exponents governing the transition at $\beta=\beta_c$ (i.e. at the tricritical point) are distinct from those at $\beta < \beta_c$. 

After characterizing the effects of learning on the classical Ising model, we exploit a duality relation to study the effects of weak measurements on a family of topologically ordered many-body quantum states which include the toric code~\cite{Kitaev2003}. Coherent superpositions of locally indistinguishable toric code states, belonging to different topological sectors, define robust quantum memories. These memories are well known to be robust to weak decoherence~\cite{Preskill2002,Fan24coherinfo,Grover24selfdual}, and therefore also to the weak measurements that we consider~\cite{teleportcode,Chen24nishimori, Puetz24}.
A basic question is whether the quantum memory remains robust when it is constructed from initial states that are far from the stabilizer limit of the toric code. 

To address this question we study the effects of weak measurements on quantum memories defined by deformed toric code wavefunctions, where now $\beta \geq 0$ parameterizes the wavefunction deformation. In the absence of measurement, these deformations drive a quantum phase transition~\cite{Fradkin04RK} from topologically ordered ($\beta<\beta_c$) to trivial ($\beta > \beta_c$) wavefunctions. Remarkably, we find that the quantum memory is robust even when constructed from states that are {\it arbitrarily close} to the quantum phase transition. Using the duality to the classical learning problem this robustness follows from the fact that, at the thermal phase transition of the Ising model, the effects of measurements are (marginally) irrelevant, i.e. that $\gamma_T>0$. We illustrate this behavior, and its manifestations in the two problems, in Fig.~\ref{fig:phasediagram}. 
We note in passing that a similar structure of the phase diagram is found for a measurement version of the all-to-all interacting Ising ferromagnet, which is a 1-replica generalization of the Sherrington-Kirkpatrick model. In particular, there is also a tricritical point at finite $\gamma_c(\beta_c)$ where the three phases meet, see End Matter. %Supplementary Material (SM)~\cite{supplement}. 

The $\beta=0$ limit of our quantum problem~\cite{teleportcode} is dual to those studied in Refs.~\cite{NishimoriCat,JYLee, Puetz24}. The quantum states we study are of Rokhsar-Kivelson form~\cite{Henley04RK,Fradkin04RK}, whose measurement-induced critical points are described by non-unitary CFTs~\cite{Ludwig05c0log,Wang25selfdual}, as are those of monitored quantum dynamics~\cite{Ludwig2020, Buechler20, Chen20measurefreefermion, Ludwig21conformal, Pixley22MIPTCFT, Vasseur24boundarytrsf}; for related works on measurement effects on quantum ground states and learnability, see Refs.~\cite{Garratt22,Lee23decoher,Altman23errorfielddouble,Garratt23measureising,Alicea23measureising,Alicea24teleportation,jian23measureising,Ludwig24measurecritical,liu2024boundary,LuitzGarratt2024MeasStatesHigherDim,hoshino2024entanglement,tang2024critical,Vasseur22learnability, Bao20learning, Potter24learnability, Ippoliti24learnability, Vasseur25mixedlearning}.
The learning transitions can be realized as measurement-induced purification transitions~\cite{Gullans20scalabledecoder, choi2020quantum,Huse2020qec} in quantum hardware, see the Supplemental Material for a discussion.

%%%%%%%%%%%%%%%%%%%%%%%%%%%%%%%%%%%%%%%%%%%%%%%%%%%%%%%%%%%%%%%%%%%
{\it Classical learning problem}.--
%%%%%%%%%%%%%%%%%%%%%%%%%%%%%%%%%%%%%%%%%%%%%%%%%%%%%%%%%%%%%%%%%%%
For a classical configuration of spins $\sigma_j = \pm 1$ on the vertices $j$ of a square lattice, the Gibbs distribution is
\begin{equation}
    P(\bm{\sigma}) = e^{-\beta E(\bm{\sigma})}/\mathcal{Z} \ ,\quad \mathcal{Z}= \sum_{\bm{\sigma}} e^{-\beta E(\bm{\sigma})} \ ,
\label{eq:Psigma}
\end{equation}
where $E(\bm{\sigma})=-\sum_{\langle i j \rangle}\sigma_i \sigma_j$ denotes the Ising energy of a full configuration $\bm{\sigma}$ of spins $\sigma_j$. The Gibbs distribution maximizes the entropy subject to the constraint of fixed $E = \sum_{\bm{\sigma}} P(\bm{\sigma})E(\bm{\sigma})$~\cite{Jaynes1957, Jaynes1957b}. The ensemble undergoes a phase transition from the paramagnet phase at high temperature to the ferromagnetic phase at low temperature across the 2D Ising critical point at $\beta_c=\ln\sqrt{1+\sqrt{2}}$. 

We will study how this distribution is modified when we `learn' the local correlation $\sigma_i \sigma_j$ with $i$ and $j$ nearest neighbors on the square lattice. Here `learning' means that we extract a random binary variable $s_{ij}=\pm 1$ that is correlated with $\sigma_i \sigma_j$, following a distribution $P(s_{ij}|\sigma_i\sigma_j) = (1+\gamma s_{ij}\sigma_i \sigma_j )/2$ where $\gamma \in [0,1]$ controls the learning precision and hence the local entropy reduction, having an interpretation as a `measurement strength'. For large $\gamma$ we learn $\sigma_i \sigma_j$ perfectly, maximally reducing the local entropy, while for $\gamma=0$ we learn nothing. The full set $\bm{s}$ of parameters $s_{ij}$ (one for each edge of the square lattice) then has a conditional {\it product} distribution
\begin{align}
    P(\bm{s}|\bm{\sigma}) = \prod_{\langle i j \rangle} \frac{1+ \gamma s_{ij}\sigma_i \sigma_j}{2} \,,
    \label{eq:Pssigma}
\end{align}
and unconditional distribution $P(\bm{s})=\sum_{\bm{\sigma}} P(\bm{s}|\bm{\sigma})P(\bm{\sigma})$. 
Having learned $\bm{s}$, the distribution of $\bm{\sigma}$ is updated according to Bayes' theorem, see Fig.~\ref{fig:phasediagram}(a). The posterior probability distribution is
\begin{align}
    P(\bm{\sigma}|\bm{s}) = \frac{1}{\mathcal{Z}(\bm{s})}\exp\left(\sum_{\langle ij\rangle} \tilde{\gamma}  s_{ij} \sigma_i \sigma_j - \beta E(\sigma)\right)\ ,
    \label{eq:condprob}
\end{align}
with $\tilde{\gamma}\equiv \tanh^{-1}(\gamma)$. 
Here $\mathcal{Z}(s)$ is the normalization constant that is proportional to the probability $P(\bm{s})=\mathcal{Z}(\bm{s})/\sum_{\bm{s}}\mathcal{Z}(\bm{s})$. 
Note that, for $\beta\neq 0$, this is a {\it gauge asymmetric}~\cite{nishimori2001statistical} random bond Ising model. 

We can develop some intuition by considering the limit $\gamma = 1$, where $s_{ij}=\sigma_i \sigma_j$. In this case, $P(\bm{\sigma}|\bm{s})=1/2$ for the two spin configurations $\bm{\sigma}$ that are consistent with the observed $\bm{s}$ (and that are related to each other by $\bm{\sigma}\to -\bm{\sigma}$), while $P(\bm{\sigma}|\bm{s})=0$ for all other configurations. Therefore, the conditional correlation function $\langle\sigma_i \sigma_j\rangle_{\bm{s}} \equiv \sum_{\bm{\sigma}}P(\bm{\sigma}|\bm{s})\sigma_i \sigma_j$ is simply $\langle\sigma_i \sigma_j\rangle_{\bm{s}}=\pm 1$. 

More generally, the correlations described by $P(\bm{\sigma}|\bm{s})$ depend sensitively on $\bm{s}$. We denote averages over the ensemble of $\bm{s}$ with respect to $P(\bm{s})$ by $[\cdots]$. For $\gamma = 1$ we have $[\langle \sigma_i \sigma_j \rangle^2_{\bm{s}}]=1$. Note that, on the other hand, $[\langle \sigma_i \sigma_j \rangle_{\bm{s}}]=\langle \sigma_i \sigma_j \rangle$ for all $\gamma$, where $\langle \cdots \rangle$ is an (unconditional) average with respect to $P(\bm{\sigma})$.

%%%%%%%%%%%%%%%%%%%%%%%%%%%%%%%%%%%%%%%%%%%%%%%%%%%%%%%%%%%%%%%%%%%
\begin{figure}[t!] 
    \includegraphics[width=\columnwidth]{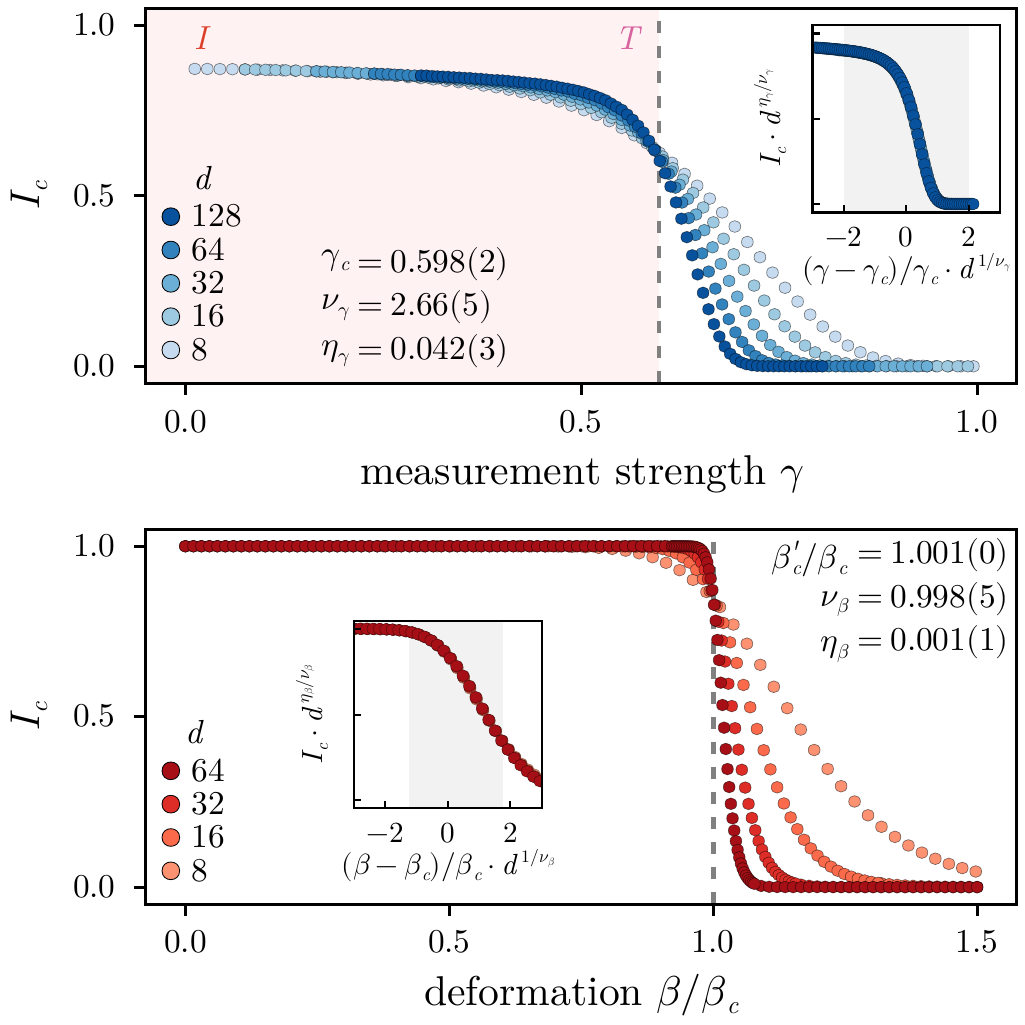} 
    \caption{
    {\bf Finite threshold of the critical state}. 
    Shown is the domain wall entropy $I_c$ (equivalent to the coherent information), which signals the phase transition.
    %(a) 
    $I_c$ for increasing measurement strength at critical Ising temperature $\beta_c=\ln\sqrt{1+\sqrt{2}}$. 
    	The pink shade highlights the critical region with finite fraction of $I_c$, between the Ising critical point ``I" and the tricritical point ``T", at
    	$\gamma_T\approx 0.598(2)$. It is found that $I_c$ yields a nonzero scaling dimension $\eta/\nu$ at the tricritical point, 
	in contrast to scaling invariant at the Ising critical point.  
    % (b) $I_c$ along tuning temperature at a chosen finite measurement strength below the threshold at $ \gamma = 0.3 <\gamma_T$. 
    %
    Numerical computation is performed for a square stripe of length $d$, 
    which is dual to a deformed surface code of code distance $d$. 
    The error bars of standard deviation are smaller than the markers. 
     }
    \label{fig:Ic}
 \end{figure}
 %%%%%%%%%%%%%%%%%%%%%%%%%%%%%%%%%%%%%%%%%%%%%%%%%%%%%%%%%%%%%%%%%%%

%%%%%%%%%%%%%%%%%%%%%%%%%%%%%%%%%%%%%%%%%%%%%%%%%%%%%%%%%%%%%%%%%%%
{\it Nonlinear probes of the effects of learning.}-- 
%%%%%%%%%%%%%%%%%%%%%%%%%%%%%%%%%%%%%%%%%%%%%%%%%%%%%%%%%%%%%%%%%%%
A general feature of learning problems is that averages over linear functions of conditional correlations are insensitive to the effects of learning, i.e.\ $[\langle \cdots\rangle_{\bm{s}}]=\langle \cdots\rangle$, where `$\cdots$' is a function of $\bm{\sigma}$. Since the ferromagnetic (FM) phase is defined by correlations $\langle \sigma_i \sigma_j\rangle$ that are finite at large separations between $i$ and $j$, this property implies that the transition out of the FM phase is at a temperature $\beta_c^{-1}$ that is independent of $\gamma$, as shown in Fig.~\ref{fig:phasediagram}.

To probe the effects of learning at the level of the ensemble of $\bm{s}$, it is necessary to average {\it nonlinear} functions of the conditional correlations. In the paramagnetic (PM) phase (at small $\beta$ and small $\gamma$) the Edwards-Anderson correlation function $[\langle \sigma_i \sigma_j \rangle_{\bm{s}}^2]$ vanishes at large separation between $i$ and $j$, while in the `spin glass' (SG) phase (at small $\beta$ and large $\gamma$) this correlation function is finite at large separations. For all $\beta \leq \beta_c$, we will see that there is a critical measurement strength $\gamma_c=\gamma_c(\beta)$ corresponding to the transition from PM to SG. At $\beta=0$, where the model is on the Nishimori line~\cite{Nishimori1981,Iba1999} with gauge symmetrized disorder~\cite{NishimoriCat}, the critical measurement strength $\gamma_c(\beta=0)$ has previously been identified as $\gamma_c(0)\approx 0.782$, which corresponds to the known critical disorder probability $10.9\%$~\cite{Pujol2001,Chalker2002, Ohzeki15RBIM, Deng25RBIM}.

To determine $\gamma_c(\beta)$ and the critical exponents, we study an averaged domain wall entropy, equivalent to the coherent information in the deformed toric code. 
To define the domain wall entropy, we consider the conditional boundary-to-boundary correlation $C_{\bm{s}} \equiv \langle \sigma_L \sigma_R\rangle_{\bm{s}}$ on a $d \times d$ stripe geometry (see the Supplemental Material for a detailed description and graphical representation).
Since the probability for the absence/presence of a domain wall is $(1 \pm C_{\bm{s}})/2$, the conditional domain wall entropy is
\begin{align}
    I_{\bm{s}} \equiv -\frac{1+C_{\bm{s}}}{2}\log_2 \frac{1+C_{\bm{s}}}{2} -\frac{1-C_{\bm{s}}}{2}\log_2 \frac{1-C_{\bm{s}}}{2} \,,
    \label{eq:coherentinformation}
\end{align}
and the average of the domain wall entropy over $\bm{s}$ is denoted $I_c\equiv [I_{\bm{s}}]$. 
In the PM phase, domain wall fluctuations destroy long-range conditional correlations, so that $I_c \to 1$ at large linear dimension $d$. In the SG phase we instead expect $I_c \to 0$ at large $d$. On the other hand, in the low temperature phase $\beta > \beta_c$, the prior distribution is already a long-range ordered ferromagnetic (FM) phase, thus for all $\gamma$ we expect vanishing domain wall entropy $I_c \to 0$ for large $L$.

We now numerically extract $\gamma_c(\beta)$ and critical exponents $\nu_\gamma(\beta)$ and $\nu_\beta(\gamma)$, governing the divergence of the correlation length when varying $\gamma$ or $\beta$.
Our numerical calculations are carried out as follows: First we sample configurations of $\sigma_i\sigma_j$, and hence $s_{ij}$ using Eq.~\eqref{eq:Pssigma}, from a standard Monte Carlo simulation of the classical Ising model, and we then use tensor network methods~\cite{teleportcode}
to evaluate conditional observables such as $I_{\bm{s}}$.
By averaging $I_{\bm{s}}$ over observed $\bm{s}$ we arrive at the estimates for $I_c$ in Fig.~\ref{fig:Ic}.

In Fig.~\ref{fig:Ic} we vary $\gamma$ at $\beta=\beta_c$. For small $\gamma$ the data shows that $I_c$ is approximately independent of $d$, whereas for large $\gamma$ the domain wall entropy is suppressed upon increasing $d$. This suggests that $\gamma_T \equiv \gamma_c(\beta_c)$ is finite. 
Indeed, we can collapse $I_c$ for various $d$ and $\gamma$ as shown in the inset, with our results indicating $\gamma_T \approx 0.598(2)$ and a correlation length exponent for the learning transition of $\nu_{\gamma}(\beta_c) = 2.658(54)$. Strikingly, the exponent $\nu_{\gamma}(\beta_c)$ is distinct from the $\beta=0$ value $\nu_{\gamma}(0) = 1.564(46)$. Moreover, $\nu_{\gamma}(\beta)$ varies only weakly with $\beta$ in the regime where $\beta$ is well below $\beta_c$, sharply increasing to $\nu_{\gamma}(\beta_c)$ over a narrow interval in $\beta$. These results suggest that the learning transition at $\beta=\beta_c$, which sits at a tricritical point, is in a different universality class to the transition at $\beta < \beta_c$.  

The numerically extracted critical exponents, see End Matter, indicate renormalization group (RG) flows from the new tricritical point to $\beta=0$ Nishimori criticality (for $\beta < \beta_c$ and $\gamma=\gamma_c(\beta)$) and to $\gamma=0$ Ising criticality (for $\beta=\beta_c$ and $\gamma < \gamma_T$). Note that the tricritical point must be unstable to a small $\beta < \beta_c$ since the RG flow of the temperature variable affects unconditional correlations such as $[\langle \sigma_i \sigma_j \rangle_{\bm{s}}]$. These correlations are independent of $\gamma$, so the RG flow of temperature must be as well. We now discuss the flow to Ising criticality at $\beta=\beta_c$ and $\gamma <\gamma_T$.

{\it Marginal irrelevance of weak measurement}.-- To show that $\gamma_T$ is finite we use a replica field theory. The replica trick involves writing nonlinear correlation functions as, e.g.,
\begin{align}
     [\langle \sigma_i \sigma_j \rangle_{\bm{s}}^2] = \lim_{n \to 1}\frac{\sum_{\bm{s}} P^n(\bm{s}) \big[ \sum_{\bm{\sigma}} P(\bm{\sigma}|\bm{s}) \sigma_i \sigma_j\big]^2}{\sum_{\bm{s}} P^n(\bm{s})} \label{eq:replicas}
\end{align}
for bulk correlations (and similarly for $I_c$), where one calculates for integer $n \geq 2$ and continues to $n=1$.

From Eq.~\eqref{eq:replicas} it is clear that, in the replica theory, the object $\sum_{\bm{s}}P^n(\bm{s})$ plays the role of the partition function. Using the  expression for $P(\bm{s})$ as a sum over spin configurations $\bm{\sigma}$, we can sum over $\bm{s}$ such that $\sum_{\bm{s}}P^n(\bm{s})$ becomes a partition function for $n$ classical Ising models, having degrees of freedom $\sigma^a_i$ with $a=1,\ldots,n$ replica indices, and `inter-replica' couplings $\sigma^a_i \sigma^a_j \sigma^b_i \sigma^b_j$ on all nearest-neighbor bonds $\langle ij \rangle$. 

Since for $\beta=\beta_c$ and $\gamma=0$ the long-distance behavior of correlation functions can be calculated using the Ising CFT, it is natural to expect that the effects of a small nonzero $\gamma$ can be determined using this framework. We therefore consider an $n$-replica Ising CFT with local perturbations that couple even parity fields (since $\sigma_i \sigma_j$ is even under the parity transformation $\bm{\sigma}\to - \bm{\sigma}$). The action for this theory is
\begin{align}
    \mathcal{S}_n = \sum_{a=1}^n \mathcal{S}_{\text{I}}^a - g \sum_{a,b=1 \atop a \neq b}^{n} \int d^2 x \, \varepsilon^a(\vec{x}) \varepsilon^b(\vec{x}) \,, \label{eq:action}
\end{align}
where $\mathcal{S}_{\text{I}}^a$ describes the $a^{\text{th}}$ replica of the unperturbed Ising CFT, with $a=1,\ldots,n$, the parameter $g \propto \tilde\gamma^2$, and $\varepsilon^a(\vec{x})$ is the energy density in replica $a$ and at position $\vec{x}=(x,y)$. The spin density is denoted by $\sigma^a(\vec{x})$. Within this theory, the Edwards-Anderson correlation function $[\langle \sigma_i \sigma_j\rangle_{\bm{s}}^2]$ is given by the $n \to 1$ limit of $\langle \sigma^1(\vec{x}_i)\sigma^1(\vec{x}_j)\sigma^2(\vec{x}_i)\sigma^2(\vec{x}_j)\rangle_{\mathcal{S}_n}$, where $\langle \cdots \rangle_{\mathcal{S}_n}$ is an average with respect to the statistical weight $e^{-\mathcal{S}_n}$, and $\vec{x}_i$ is the position of site $i$. To determine whether $[\langle \sigma_i \sigma_j\rangle_{\bm{s}}^2]$ is modified at long distances relative to the behavior $\sim 1/\sqrt{r_{ij}}$ at $g=0$, we renormalize $g$. Following the standard perturbative treatment for the replicated Ising CFT \cite{dotsenko1995renormalization} we find that, under an infinitesimal change of the lattice scale by a factor of $e^{\ell}$, and at second order in $g$,
\begin{align}
    \frac{dg}{d\ell} = 4\pi(n-2)g^2 + O(g^3) \,. \label{eq:RGflow}
\end{align}
The first-order contribution is $(2-2\Delta_{\epsilon})g$, but this vanishes for all $n$ because the scaling dimension $\Delta_{\epsilon}$ of the energy density is unity. Sending $n\to1$ we see that a small nonzero $g$ flows to zero under RG. The marginal irrelevance of the effects of measurements at $\beta=\beta_c$ indicates that $\gamma_T > 0$. This is consistent with our numerical results above.

Before taking the limit $n \to 1$ our theories for the effects of learning coincide with theories for systems with quenched bond disorder (see, e.g., Ref.~\cite{dotsenko1995renormalization}). The key difference is that, in that setting, the appropriate replica limit is instead $n \to 0$.

%%%%%%%%%%%%%%%%%%%%%%%%%%%%%%%%%%%%%%%%%%%%%%%%%%%%%%%%%%%%%%%%%%%
{\it Deformed toric code}.--
%%%%%%%%%%%%%%%%%%%%%%%%%%%%%%%%%%%%%%%%%%%%%%%%%%%%%%%%%%%%%%%%%%%
Here we discuss the connection between our results and the effects of weak measurements on deformed toric code wavefunctions. Ising models in the geometry discussed in connection with Fig~\ref{fig:Ic} and Eq.~\eqref{eq:coherentinformation}, with boundary conditions $\mu = \sigma_L \sigma_R = \pm 1$, are associated with RK wavefunctions $\ket{\psi_{\mu}}=\sum_{\bm{\sigma}} \sqrt{P_{\mu}(\bm{\sigma})} \ket{\bm{\sigma}}$, where $P_{\mu}(\bm{\sigma})$ is the Boltzmann weight for the bulk spins $\bm{\sigma}$ with boundary condition $\mu$. These wavefunctions are dual to the two logical states of deformed toric codes on open surfaces (surface codes), which we denote $\ket{\phi_{\mu}} \propto \exp(\frac{1}{2}\sum_{\langle ij \rangle} \beta Z_{ij})\ket{\text{toric code}_{\mu}}$. In $\ket{\phi_{\mu}}$ the qubits reside on edges $\langle ij\rangle$ of the square lattice, $Z_{ij}$ is a Pauli matrix, and $\ket{\text{toric code}_{\mu}}$ is an ideal toric code ground state with $\mu=\pm 1$ specifying the logical state. These deformed states undergo quantum phase transitions across (2+0)D conformal quantum critical points~\cite{Fradkin04RK} at $\beta_c$, see Fig.~\ref{fig:phasediagram}(b).
If, in addition, we apply weak measurement of strength $\gamma$ to every qubit of the toric code state $\ket{\phi_\mu}$, 
this results in a statistical ensemble of post-measurement states conditioned on the measurement outcomes $\bm{s}$
 \begin{equation}
\ket{\phi_\mu(\bm{s})} \propto 
\prod_{\langle ij\rangle}[1+\tanh(\tilde{\gamma}/2)s_{ij}Z_{ij}]\ket{\phi_\mu}
\ .
 \label{eq:state}
 \end{equation}
Here the measurement probability is given by Born's rule $P(\bm{s})= \sum_\mu \braket{\phi_\mu(\bm{s})}$ which leads to the exact same result as Bayes' rule~\eqref{eq:condprob} 
in the classical inference model. 
 The coherent information~\cite{Nielsen96coherentinfo, Gullans20scalabledecoder} of the toric code~\cite{Fan24coherinfo} under such measurement channel is related to the domain wall entropy $I_c$ in \eqref{eq:coherentinformation}, by noting that now $C_{\bm{s}}=\bra{\phi_-(\bm{s})}\ket{\phi_+(\bm{s})}$ takes physical meaning as the fidelity between two topologically degenerate states~\cite{teleportcode}.
 Then it is straightforward to map the phase diagram of the classical inference model to that of the toric code in Fig.~\ref{fig:phasediagram}(b,c).
 In the SG regime, the toric code is subjected to strong measurement and dephased into a classical loop gas, which loses the quantum memory ($I_c=0$) but still retains a  classical topological memory encoded by the domain wall as a non-contractible loop. Such a phase is characterized by  order parameters $[C_{\bm{s}}]=0 \ ,\ [C_{\bm{s}}^2]\neq 0$. In the FM regime, the classical topological memory is lost, since the domain wall is exponentially suppressed dictated by $[C_{\bm{s}}]\neq 0$.
 The location of the phase boundary between the PM and the SG phases, $\gamma_c$, determines the information-theoretic threshold of the toric code states against measurement. According to the phase diagram, when one deforms the toric code away from its $\beta=0$ fixed point by increasing $\beta$, the information-theoretic threshold $\gamma_c$ decreases from $\gamma_N\approx 0.782$ (Nishimori threshold) to $\gamma_T\approx 0.598(2)$ (tricritical threshold) instead of vanishing as the critical point is approached. 
 Then it suddenly drops to zero when it becomes a trivial phase. 
In other words, the existence of a tricritical point at finite $\gamma_T$ means that coherent superpositions of deformed toric code states remain stable to measurement even when arbitrarily close to the critical point. 
Here, this is a consequence of the marginal irrelevance of $\mathbbm{Z}_2$ symmetric measurements in the Ising CFT. 

%%%%%%%%%%%%%%%%%%%%%%%%%%%%%%%%%%%%%%%%%%%%%%%%%%%%%%%%%%%%%%%%%%%
{\it Symmetry perspective}.--
%%%%%%%%%%%%%%%%%%%%%%%%%%%%%%%%%%%%%%%%%%%%%%%%%%%%%%%%%%%%%%%%%%%
Let us close by discussing the symmetries of the three phases around the tricritical point. We will focus on the RK wavefunctions $\ket{\psi(\bm{s})} = \sum_{\bm{\sigma}} \sqrt{P(\bm{\sigma}|\bm{s})} \ket{\bm{\sigma}}$ associated with the conditional distributions $P(\bm{\sigma}|\bm{s})$; for this discussion we consider an infinite system and omit $\mu$. Ising criticality is a well-known universality class for $\mathbbm{Z}_2$ symmetry breaking in disorder-free situations. 
In the presence of disorder, Nishimori criticality -- in lieu of Ising -- has been found to be ubiquitous in $\mathbbm{Z}_2$ mixed quantum states governed by Born's rule~\cite{Preskill2002, NishimoriCat,Chen24nishimori,JYLee,Fan24coherinfo}, where it has been connected to $\mathbbm{Z}_2$ strong-to-weak spontaneous symmetry breaking~\cite{You24weaksym, Wang24strtowksym, Luo24weaksym}. 
These two universality classes naturally appear in the problem at hand, 
and are inevitably bound to join at the tricritical point we have identified, see Figs.~\ref{fig:phasediagram}(b,c). The three phases meeting at the tricritical point distinguish themselves by exhibiting strong, weak, or no $\mathbbm{Z}_2$ symmetry, in a sense we now discuss. 

The classical-quantum density matrix $\rho = \sum_{\bm{s}} P(\bm{s}) \ketbra{\bm{s}} \otimes \ketbra{\psi(\bm{s})}$~\cite{Wang25selfdual} possesses a strong $\mathbbm{Z}_2$ symmetry throughout the phase diagram. For $\beta > \beta_c$ this symmetry is spontaneously broken (FM), while at $\beta < \beta_c$ and large $\gamma$ the spin-glass correlations signal strong-to-weak symmetry breaking (SW-SSB)~\cite{Wang24strtowksym, You24weaksym, Luo24weaksym} with statistical average long-range order (see the Supplemental Material for details).
The three phases meeting at the tricritical point thus distinguish themselves as: a strongly symmetric phase (PM), a SW-SSB phase (SG), or a conventionally symmetry broken phase (FM). 
\\

\acknowledgments
%%%%%%%%%%%%%%%%%%%%%%%%%%%%%%%%%%%%%%%%%%%%%%%%%%%%%%%%%%%%%%%%%%%
{\it Note added}.-- Upon completion of this work, we became aware of a related independent study~\cite{Nahum25bayesiancriticalpoints} 
of Bayesian inference in the context of classical lattice models. Our results on the classical Ising model agree where they overlap. Additionally, recent work in Refs.~\cite{Vasseur25monitoredhydro,Lamacraft25inference} has investigated monitored classical and quantum dynamics, respectively, through the lens of statistical inference.

%%%%%%%%%%%%%%%%%%%%%%%%%%%%%%%%%%%%%%%%%%%%%%%%%%%%%%%%%%%%%%%%%%%
{\it Data availability}.-- 
The numerical data shown in the figures and the data for sweeping the phase diagram is available on Zenodo~\cite{zenodo_learning}.\\
%%%%%%%%%%%%%%%%%%%%%%%%%%%%%%%%%%%%%%%%%%%%%%%%%%%%%%%%%%%%%%%%%%%

%%%%%%%%%%%%%%%%%%%%%%%%%%%%%%%%%%%%%%%%%%%%%%%%%%%%%%%%%%%%%%%%%%%
{\it Acknowledgments}.--
%%%%%%%%%%%%%%%%%%%%%%%%%%%%%%%%%%%%%%%%%%%%%%%%%%%%%%%%%%%%%%%%%%%
The Cologne group gratefully acknowledges partial funding from the Deutsche Forschungsgemeinschaft (DFG, German Research Foundation)
under Germany's Excellence Strategy -- Cluster of Excellence Matter and Light for Quantum Computing (ML4Q) EXC 2004/1 -- 390534769 
and within the CRC network TR 183 (Project Grant No.~277101999) as part of subproject B01.
S. J. G. was supported by the Gordon and Betty Moore Foundation.
G.-Y. Z. acknowledge the support of National Natural Science Foundation of China - Young Scientists Fund (grant no.~12504181) and Start-up Fund of HKUST(GZ) (grant no.~G0101000221) and Guangdong provincial project (grant no.~2024QN11X201). 
The numerical simulations were performed on the JUWELS cluster at the Forschungszentrum Juelich.

%%%%%%%%%%%%%%%%%%%%%%%%%%%%%%%%%%%%%%%%%%%%%%%%%%%%%%%%%%%%%%%%%%%
\bibliography{measurementsbib}
%%%%%%%%%%%%%%%%%%%%%%%%%%%%%%%%%%%%%%%%%%%%%%%%%%%%%%%%%%%%%%%%%%%

%%%%%%%%%%%%%%%%%%%%%%%%%%%%%%%%%%%%%%%%%%%%%%%%%%%%%%%%%%%%%%%%%%%
\appendix
%%%%%%%%%%%%%%%%%%%%%%%%%%%%%%%%%%%%%%%%%%%%%%%%%%%%%%%%%%%%%%%%%%%
\clearpage
\section{End matter}

%%%%%%%%%%%%%%%%%%%%%%%%%%%%%%%%%%%%%%%%%%%%%%%%%%%%%%%%%%%%%%%%%%%

\begin{figure}[thb!] 
    \includegraphics[width=\columnwidth]{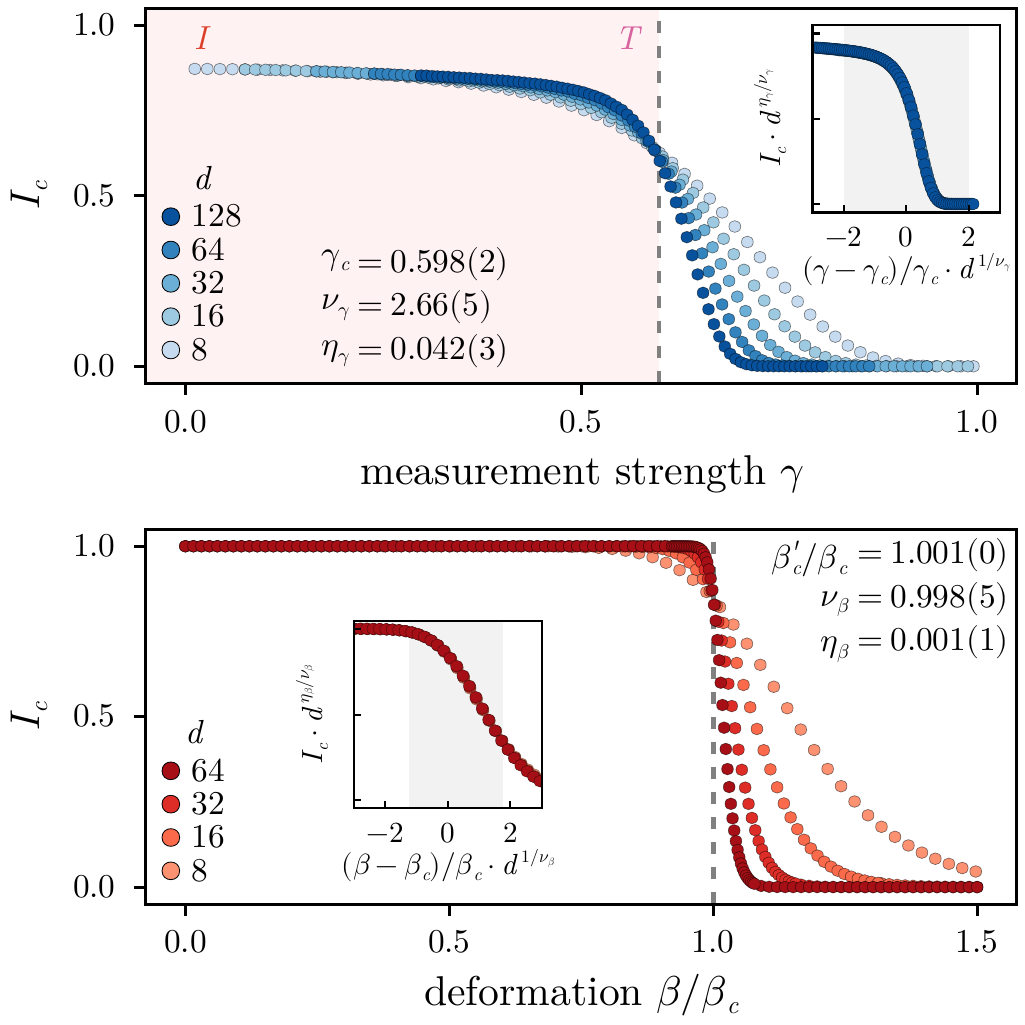} 
    \caption{
    {\bf Finite temperature transition of the weakly measured state}. 
    Shown is the domain wall entropy $I_c$ (equivalent to the coherent information), which signals the phase transition. $I_c$ along tuning temperature at a chosen finite measurement strength below the threshold at $ \gamma = 0.3 <\gamma_T$. 
     }
    \label{fig:Icb}
 \end{figure}

{\it Finite temperature transition of the weakly measured state}.-- 
In Fig.~\ref{fig:Icb} we vary $\beta$ with $\gamma$ fixed to be below the threshold $\gamma_T$. As expected, for high temperature $\beta <\beta_c$, increasing $d$ causes $I_c$ to approach unity, while for low temperature $\beta > \beta_c$ we find that $I_c$ decays rapidly with increasing $d$. The data collapse (inset) shows that away from the tricritical point the thermal correlation length exponent $\nu_{\beta}(\gamma)=0.998(5)$, consistent with the analytical value $\nu_{\beta}(0)=1$ for the 2D Ising transition.  

%%%%%%%%%%%%%%%%%%%%%%%%%%%%%%%%%%%%%%%%%%%%%%%%%%%%%%%%%%%%%%%%%%%
 %%%%%%%%%%%%%%%%%%%%%%%%%%%%%%%%%%%%%%%%%%%%%%%%%%%%%%%%%%%%%%%%%%%

%%%%%%%%%%%%%%%%%%%%%%%%%%%%%%%%%%%%%%%%%%%%%%%%%%%%%%%%%%%%%%%%%%%
\begin{figure}[b!] 
    \includegraphics[width=\columnwidth]{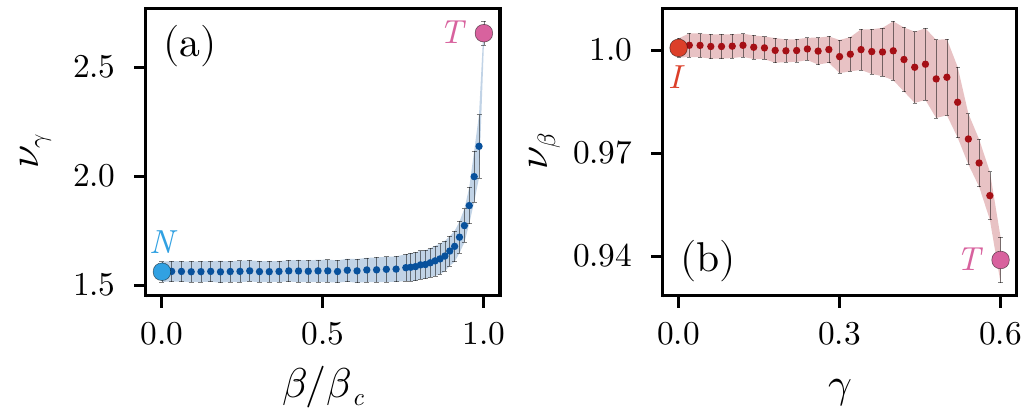} 
    \caption{
        {\bf Critical scaling and RG flow at the transition lines.} 
    (a) From Nishimori critical to the tricritical point: 
    		the critical exponent $\nu_{\gamma}$ across $\gamma_c$ defined by $\xi\propto (\gamma-\gamma_c)^{-\nu_{\gamma}}$, 
		calculated for horizontal cuts of fixed $\beta$.
    (b) From clean Ising critical to tricritical point: 
    		the critical exponent $\nu_\beta$ for $\xi\propto (\beta-\beta_c)^{-\nu_\beta}$,
            calculated for vertical cuts of fixed $\gamma$. 
    The critical exponents are extracted from the coherent information $I_c$ ($d=8, 16, 32, 64$), 
    using a scaling ansatz with two critical exponents $\nu_{\gamma}$ and $\eta$, see the Appendix.
    In the respective limits, the numerical results agree with the known exponents for Nishimori $\nu_\gamma(0)=1.564(46)$ (blue circle) 
     and Ising $\nu_\beta(0) =1$ (red circle) criticality. 
     The tricritical point has scaling exponents $\nu_\gamma(\beta_c)=2.658(54)$ and $\nu_\beta(\gamma_c)=0.9389(66)$.
    }
    \label{fig:nu}
 \end{figure}
%%%%%%%%%%%%%%%%%%%%%%%%%%%%%%%%%%%%%%%%%%%%%%%%%%%%%%%%%%%%%%%%%%%

{\it Critical exponents and RG flow}.--
In Fig.~\ref{fig:nu} we show the variation of the critical exponents along the transition lines. Figure~\ref{fig:nu}(a) shows the critical exponent $\nu_{\gamma}(\beta)$ for the learning transition as a function of $\beta$, i.e.\ moving from the Nishimori point to the tricritical point. The change in $\nu_{\gamma}(\beta)$ from $\nu_{\gamma}(0)= 1.564(46)$ to $\nu_{\gamma}(\beta_c) = 2.658(54)$ occurs over a small interval of $\beta/\beta_c$ just below unity. These results suggest that the learning transitions with $\beta < \beta_c$ are in the same universality class as the learning transition at $\beta=0$, but that the transition at $\beta=\beta_c$ is distinct. That is, we can identify an RG flow from the tricritical point at critical temperature $\beta_c$ to the Nishimori fixed point at infinite temperature $\beta=0$.
Figure~\ref{fig:nu}(b) shows the critical exponent $\nu_{\beta}(\gamma)$ for the thermal phase transition as a function of $\gamma$ (i.e.\ from the pure Ising to the tricritical transition); we again find a fairly sharp change in the exponent on approaching the tricritical point. Together, the numerically extracted critical exponents indicate a tricritical point at finite $\gamma_c(\beta_c) = 0.598(2)$ that is unstable, flowing to Ising criticality along the phase boundary with $\beta=\beta_c$ and $\gamma < \gamma_c(\beta_c)$, and to Nishimori criticality on the phase boundary with $\beta < \beta_c$ and $\gamma=\gamma_c(\beta)$, see Fig.~\ref{fig:phasediagram}(b).

%%%%%%%%%%%%%%%%%%%%%%%%%%%%%%%%%%%%%%%%%%%%%%%%%%%%%%%%%%%%%%%%%%%

{\it Exactly solved all-to-all model}.--
In this Appendix we discuss the effects of measurements on all-to-all ferromagnets. The probability to find the system in configuration $\bm{\sigma}$ is
\begin{align}
	P(\bm{\sigma}) \sim e^{\frac{\beta}{N} \sum_{i < j} \sigma_i \sigma_j},
\end{align}
where we have rescaled the coupling by $1/N$ so that the energy is extensive. Using a Gaussian measurement model, the probability density to find outcomes $\bm{s} = \{s_{ij}\}$ is
\begin{align}
	P(\bm{s}) \sim \sum_{\bm{\sigma}} e^{\frac{\beta}{N} \sum_{i < j} \sigma_i \sigma_j - \frac{\gamma^2}{2N} \sum_{i < j} [s_{ij}-\sigma_i \sigma_j]^2},
\end{align}
where the distribution is normalized as $\int_{-\infty}^{\infty} \prod_{i < j} ds_{ij} P(\bm{s})=1$. Increasing $\gamma$ corresponds to increasing the correlation between $s_{ij}$ and $\sigma_i \sigma_j$. To calculate averages of (nonlinear) correlations over $s$, we use a replica trick. The partition function for the $n$ replica theory is $\mathcal{Z}_n = \int d\bm{s} P^n(s)$ and $d\bm{s} =  \prod_{i < j} ds_{ij}$. Integrating out $\bm{s}$ we have
\begin{align}
	\mathcal{Z}_n \sim \sum_{\bm{\sigma}^1 \cdots \bm{\sigma}^n} e^{ \frac{\beta}{N}\sum_{\alpha}\sum_{j < k} \sigma^{\alpha}_j \sigma^{\alpha}_k + \frac{\gamma^2}{nN} \sum_{\alpha < \beta} \sum_{j < k} \sigma^{\alpha}_j \sigma^{\alpha}_k \sigma^{\beta}_j \sigma^{\beta}_k},
\end{align}
where the superscript $\alpha=1,\cdots, n$ is a replica index. We can decouple the interactions between different spins $j$ in two steps. First we introduce variables $m^{\alpha}$ and $q^{\alpha \beta}$ (for $\alpha < \beta$):
\begin{align}
	\mathcal{Z}_n \sim \int &\prod_{\alpha} dm^{\alpha}\prod_{\beta > \alpha} dq^{\alpha \beta} e^{  -\frac{1}{2}N\beta \sum_{\alpha}(m^{\alpha})^2 + \beta m^{\alpha} \sum_j \sigma^{\alpha}_j}\\
    \times &e^{- \frac{1}{2n }N \gamma^2 \sum_{\alpha < \beta} (q^{\alpha \beta})^2 + \frac{\gamma^2}{n} \sum_{\alpha < \beta} q^{\alpha \beta} \sum_j \sigma^{\alpha}_j \sigma^{\beta}_j}.\notag
\end{align}
At large $N$ we use a saddle point approximation: $q^{\alpha \beta}=q$ and $m^{\alpha}=1$. Then $\sum_{\alpha < \beta}  q^{\alpha \beta} \sigma^{\alpha}_j \sigma^{\beta}_j = \frac{1}{2}q [ (\sum_{\alpha} \sigma^{\alpha}_j)^2-n]$, and we can absorb the term quadratic in spins using
\begin{align}
	&\sum_{\sigma_j^1\cdots \sigma_j^n} e^{\beta m \sum_{\alpha}\sigma^{\alpha}_j - \frac{\gamma^2 q}{2} + \frac{\gamma^2 q}{2n} (\sum_{\alpha} \sigma^{\alpha}_j)^2       } \notag \\&\sim \int dz \, e^{\beta m \sum_{\alpha}\sigma^{\alpha}_j - \frac{\gamma^2 q}{2} -\frac{1}{2}z^2 + z\gamma (q/n)^{1/2}\sum_{\alpha}\sigma^{\alpha}_j} \\&= 2^n e^{-\frac{\gamma^2 q}{2}}\int dz e^{-\frac{1}{2}z^2}\cosh^n\big[ \beta m + \gamma (q/n)^{1/2} z\big].\notag
\end{align}
In our saddle point approximation, we then have $\mathcal{Z}_n \sim \int dm dq e^{-N f_n[m,q]}$ with free energy density
\begin{align}
	f_n[m,q] &= \frac{1}{4}(n-1)\gamma^2 q^2 + \frac{1}{2}n \beta m^2 + \frac{1}{2}\gamma^2 q \\&- \ln \int dz \, e^{-z^2/2} \cosh^n\big[ \beta m + \gamma(q/n)^{1/2} z\big].\notag
\end{align}
To identify the PM-SG boundary we set $m=0$ and expand the free energy density in powers of $q$:
\begin{align}
	f_n[0,q] = \frac{1}{4}(n-1)( 1 - \gamma^2/n)\gamma^2q^2 + O(q^3).
\end{align}
For $n=1+\epsilon$ the prefactor of $q^2$ changes sign at $\gamma^2=1+\epsilon$, indicating that for $\epsilon \to 0$ the phase boundary is at $\gamma=1$ independent of $\beta$. To identify the PM-FM boundary we minimize $f_n[m,q]$ with respect to $m$. The result, for $n \to 1$, is
\begin{align}
	m = \frac{\int dz \, e^{-z^2/2} \sinh[\beta m + \gamma q^{1/2}z]}{\int dz \, e^{-z^2/2} \cosh[\beta m + \gamma q^{1/2}z]}.
\end{align}
Expanding the right-hand side in powers of $\gamma$, it can be verified that this self-consistency equation is $m = \tanh[\beta m]$ at all orders. As required, the PM-FM boundary (as well as the SG-FM boundary) is independent of $\gamma$.

%%%%%%%%%%%%%%%%%%%%%%%%%%%%%%%%%%%%%%%%%%%%%%%%%%%%%%%%%%%%%%%%%%%
\begin{figure}[tb!]
    \includegraphics[width=.9\columnwidth]{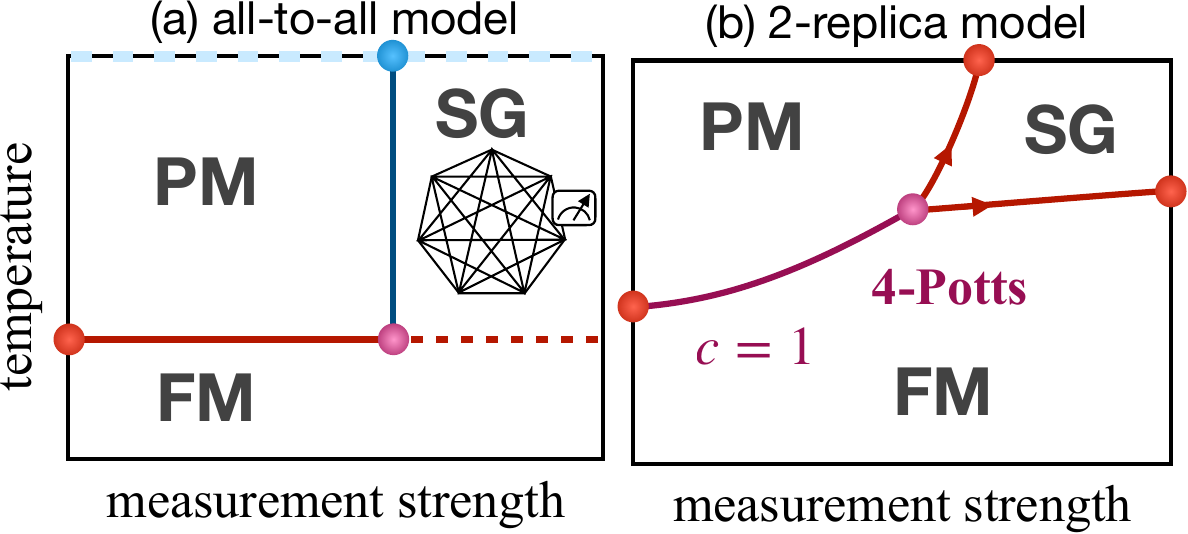}
    \caption{
        {\bf Exactly solved models}.
        (a) Phase diagram of measured all-to-all interacting Ising model, exactly solved by mean field theory akin to the Sherrington-Kirkpatrick model.
        (b) 2-replica variant of the model, by mapping the 2-replicated partition function $P(\bm{s})^2$ to the Ashkin-Teller model.
        }
    \label{fig:replicas}
\end{figure}
%%%%%%%%%%%%%%%%%%%%%%%%%%%%%%%%%%%%%%%%%%%%%%%%%%%%%%%%%%%%%%%%%%%

%%%%%%%%%%%%%%%%%%%%%%%%%%%%%%%%%%%%%%%%%%%%%%%%%%%%%%%%%%%%%%%%%%%
{\it Exactly solved 2-replica model}.--
%%%%%%%%%%%%%%%%%%%%%%%%%%%%%%%%%%%%%%%%%%%%%%%%%%%%%%%%%%%%%%%%%%%
Although it is more natural to study conditional expectation values $\langle\cdots\rangle_{\bm{s}}$ weighted according to the distribution $P(\bm{s})$, for example $\sum_{\bm{s}} P(\bm{s})\langle\sigma_i \sigma_j\rangle_{\bm{s}}^2$, we can develop more insight into the full ensemble of conditional states by weighting averages by powers of $P(\bm{s})$. For example, we can ask about the behavior of $\sum_{\bm{s}}P^n(\bm{s})\langle\sigma_i \sigma_j\rangle_{\bm{s}}^2/\sum_{\bm{s}}P^n(\bm{s})$ with separation as $\beta$ and $\gamma$ are varied. For $n > 1$ such averages bias the distribution towards more likely outcomes $\bm{s}$. Note also that, through Eq.~\eqref{eq:replicas}, we can understand these theories as corresponding to a general number $n$ of replicas. 

Here, a key advantage of considering such objects is that for $n=2$ we can solve the resulting statistical mechanics model exactly. This is because the $n=2$ theory is simply the {\it Ashkin-Teller model}. Denoting $\tilde{\gamma}=\tanh^{-1}(\gamma)$ we have
\begin{equation}
\begin{split}
\sum_{\bm{s}} P^2(\bm{s}) &\propto \sum_{\bm{s}, \bm{\sigma}, \bm{\tau}} e^{\sum_{\langle ij\rangle} (\beta + \tilde{\gamma} s_{ij})(\sigma_i\sigma_j+\tau_i\tau_j)} \\
& \propto \sum_{\bm{\sigma}, \bm{\tau}} e^{\sum_{\langle ij\rangle}  \beta(\sigma_i\sigma_j+\tau_i\tau_j) + \tanh^{-1}(\gamma^2) \sigma_i\sigma_j\tau_i\tau_j}  \ .
\end{split}
\end{equation}
Here we use the notation $\sigma_j=\sigma^1_j$,  $\tau=\tau^2_j$ for the spins in the two replicas $a=1,2$. The analogue of the spin glass order parameter in this limit is represented by the linear observable $\langle \sigma \tau\rangle$ weighted according the $n=2$ partition function $\sum_{\bm{s}}P^2(\bm{s})$.

Note that the solution of the Ashkin-Teller (AT) model provides us with predictions for the behavior of Rényi-2 fidelity correlator~\cite{You24weaksym, Wang24strtowksym} in the classical-quantum Rokhsar-Kivelson state $\rho = \sum_{\bm{s}}P(\bm{s})\ket{\bm{s}}\bra{\bm{s}}\otimes \ket{\psi(\bm{s})}\bra{\psi(\bm{s})}$. This correlator can be used to detect SW-SSB in the absence of conventional SSB, and is defined by
\begin{equation}
\mathcal{F}_2 (\rho, Z_i Z_j \rho Z_j Z_i)= \frac{\text{tr}(Z_i Z_j\rho Z_j Z_i \rho)}{\text{tr} \rho^2}.
\label{eq:fidelity2rep}
\end{equation}
Inserting our expression for $\rho$ it is clear that
\begin{equation}
\mathcal{F}_2(\rho, Z_i Z_j \rho Z_j Z_i) = \langle \sigma_i\tau_i \sigma_j\tau_j \rangle_{\text{AT}}.
\end{equation}
For comparison, conventional SSB is detected by 
\begin{equation}
\mathcal{F}_2(\rho, Z_i Z_j \rho ) = \frac{\sum_{\bm{s}} P(\bm{s})^2 \langle Z_i Z_j\rangle}{\sum_{\bm{s}} P(\bm{s})^2} = \langle \sigma_i\sigma_j\rangle_{\text{AT}} \ .
\end{equation}
Thus the ``partial order" in the Ashkin-Teller model i.e.\ $\langle \sigma\tau\rangle\neq 0$ is equivalent to replica correlation and, in our case, corresponds `spin glass' order in the $n=2$ theory. 

The phase diagram of the Ashkin-Teller model is schematically shown in Fig.~\ref{fig:replicas}b. A qualitative distinction with the 1-replica case is that the ferromagnetic phase is enhanced by the measurement, which raises the critical temperature. The transition line between ferromagnetic and paramagnetic phases is pinpointed exactly by the Kramers Wannier self-duality $\sinh(2\beta)=\exp(-2\tanh^{-1}\gamma^2)$, which exhibits continuously varying critical exponents and are described by the compactified boson conformal field theory (CFT) with central charge $c=1$. This $c=1$ critical line starts from the $\gamma=0,\ \beta=\beta_c$ point, which is described by two decoupled Ising CFTs with $c=1/2$ respectively. The self-dual critical line terminates at the $S_4$ symmetric Potts point $\beta_p = \tanh^{-1}\gamma_p^2=\tanh^{-1}(1/2)/2=0.2746...$, namely, $\gamma_p=0.5176...$. Beyond this point, it splits into two Ising critical lines sandwiching a phase with the so-called ``partial order'' $\langle \sigma_i\tau_i\sigma_j\tau_j\rangle_{\rm AT}\neq 0$ in the context of the Ashkin-Teller model. Put in our context, this corresponds to the ``spin glass'' phase in the 2-replica theory. 

%%%%%%%%%%%%%%%%%%%%%%%%%%%%%%%%%%%%%%%%%%%%%%%%%%%%%%%%%%%%%%%%%%%
\clearpage
\begin{center}
{\Large Supplemental Material}
\end{center}
%%%%%%%%%%%%%%%%%%%%%%%%%%%%%%%%%%%%%%%%%%%%%%%%%%%%%%%%%%%%%%%%%%%

%%%%%%%%%%%%%%%%%%%%%%%%%%%%%%%%%%%%%%%%%%%%%%%%%%%%%%%%%%%%%%%%%%%
\section{Symmetry perspective: detailed derivation}
%%%%%%%%%%%%%%%%%%%%%%%%%%%%%%%%%%%%%%%%%%%%%%%%%%%%%%%%%%%%%%%%%%%

The whole statistical ensemble of pure states can be compactly described by a block-diagonal classical-quantum density matrix~\cite{Wang25selfdual} of the form
 $
     \rho = \sum_{\bm{s}} P(\bm{s})  \ketbra{\bm{s}} \otimes \ketbra{\psi(\bm{s})}
 $,
 where $\ketbra{\bm{s}}$ is the state of the register that records the measurement outcomes, and $\ket{\psi(\bm{s})} = \sum_{\bm{\sigma}} \sqrt{P(\bm{\sigma}|\bm{s})} \ket{\bm{\sigma}}$ are the Rokhsar-Kivelson wavefunctions associated with the conditional distributions $P(\bm{\sigma}|\bm{s})$.
It is then easy to check that the mixed quantum state $\rho$ possesses a strong (exact) $\mathbbm{Z}_2$ symmetry
$
\left(\prod_j X_j\right) \rho = \rho = \rho \left(\prod_j X_j\right)
$ throughout the phase diagram. For $\beta > \beta_c$ this symmetry is spontaneously broken to null symmetry resulting in the ferromagnetic phase. 
For $\beta < \beta_c$ there is no exact long-range order since the ferromagnetic correlation vanishes. 
Nonetheless, at large $\gamma$  
we can relate the spin-glass correlations to the mixed-state fidelity correlator~\cite{Wang24strtowksym, You24weaksym}
$
\mathcal{F}(\rho, Z_iZ_j\rho Z_jZ_i)
= [| \langle \sigma_i \sigma_j \rangle_{\bm{s}} | ]
$, indicating the existence of a strong-to-weak symmetry breaking (SW-SSB)~\cite{Wang24strtowksym, You24weaksym} phase with {\it statistical average} long-range order.

To conclude, the relevant symmetry-preserving perturbations out of the tricritical point could lead to three distinct phases: a strongly symmetric phase (PM), a strong-to-weak symmetry broken phase (SG), or a conventional symmetry broken phase (FM). 

%%%%%%%%%%%%%%%%%%%%%%%%%%%%%%%%%%%%%%%%%%%%%%%%%%%%%%%%%%%%%%%%%%%
\section{Realization on quantum devices}
%%%%%%%%%%%%%%%%%%%%%%%%%%%%%%%%%%%%%%%%%%%%%%%%%%%%%%%%%%%%%%%%%%%

The phenomena discussed in the main text can be realized as measurement-induced purification transitions~\cite{Gullans20scalabledecoder, choi2020quantum,Huse2020qec} in quantum hardware. If a reference qubit is initially entangled with a (deformed) surface code, and the physical qubits of the surface code are weakly measured, then for $\gamma > \gamma_c$ the reference qubit will be purified. An order parameter for this transition is the measurement-averaged von Neumann entropy of the reference qubit. Although the direct determination of this order parameter suffers from an exponential post-selection problem, it can be upper bounded using cross-correlations between experimental data and numerical estimates for the effects of measurements \cite{garratt2024probing,mcginley2024postselection, Garratt25measure, Eckstein25learning}.

%%%%%%%%%%%%%%%%%%%%%%%%%%%%%%%%%%%%%%%%%%%%%%%%%%%%%%%%%%%%%%%%%%%
\section{Numerical method: sampling and random tensor network}
\begin{figure}[h!]
    \includegraphics[width=\columnwidth]{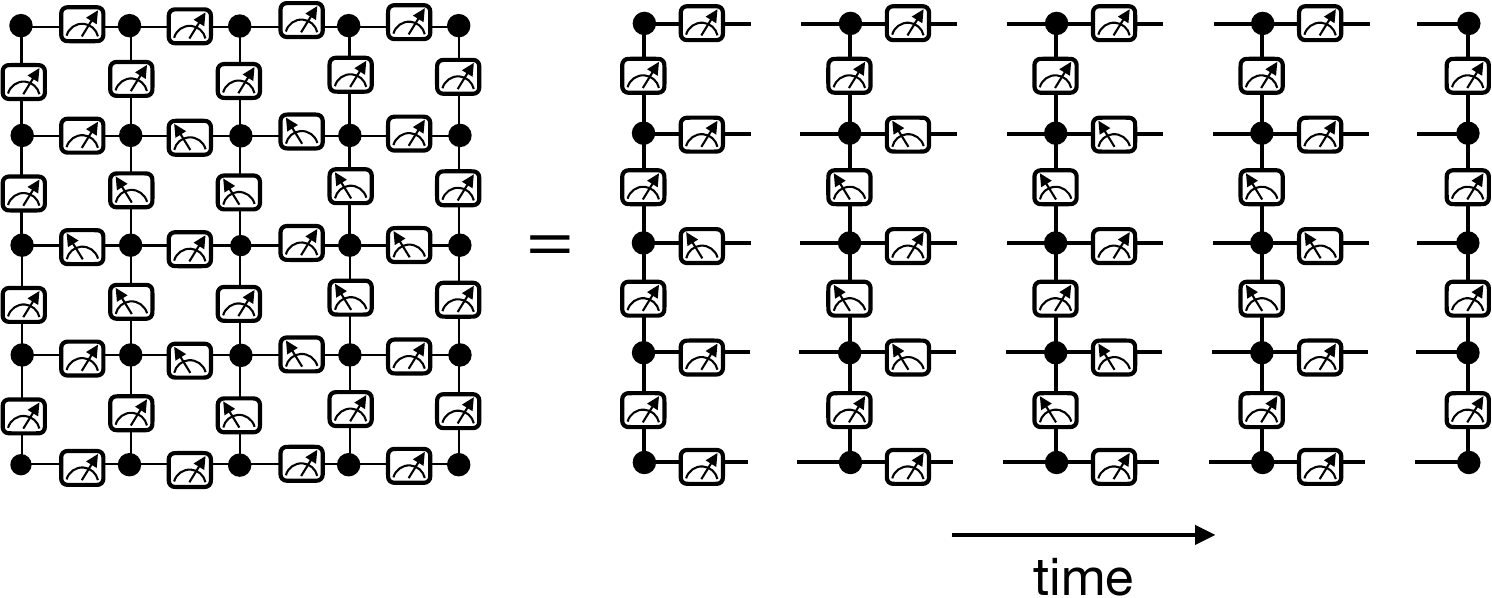}
    \caption{
        {\bf Random tensor network for the measurement problems} can be decomposed into slices of random matrix product operator as transfer matrices. A (1+1)D matrix product state evolves from left to right by the transfer matrices.
    }
    \label{fig:tensor_network}
\end{figure}

%%%%%%%%%%%%%%%%%%%%%%%%%%%%%%%%%%%%%%%%%%%%%%%%%%%%%%%%%%%%%%%%%%%
In our numerical simulation, we first generate the samples $\bm{s}$ for the precise learning limit $\gamma=1$. This can be done by  standard Monte Carlo method sampling of Ising configurations $\bm{\sigma}$ with the Boltzmann weight $P(\bm{\sigma})$. We can then convert the Ising configurations to domain wall configurations via the duality transformation: $\sigma_{i}\sigma_j = s_{ij}$, for any bond $\langle ij\rangle$. When the measurement strength is decreased to $\gamma<1$, we randomly flip the sign of each bond variable $s_{ij}$ with a probability $(1-\gamma)/2$, according to $P(s_{ij}|\sigma_i\sigma_j)=(1+\gamma s_{ij}\sigma_i\sigma_j)/2$. In this way we obtain the samples $\bm{s}$ for arbitrary measurement strength $\gamma$. Then we use the samples $\bm{s}$ to construct the random tensor network for the partition function $\mathcal{Z}(\bm{s})$. As shown in Fig.~\ref{fig:tensor_network}, we can contract out the tensor network by performing a (1+1)D matrix product state evolution from left to right. 
Note that this single layer Ising tensor network can also be fermionized into a (1+1)D monitored Gaussian fermion chain, or a 2D Chalker-Coddington network model~\cite{Chalker2002,Wang25selfdual} for efficient simulation, but the tensor network contraction can be applied to more generic statistical model that cannot be mapped to free fermion models.

%%%%%%%%%%%%%%%%%%%%%%%%%%%%%%%%%%%%%%%%%%%%%%%%%%%%%%%%%%%%%%%%%%%
%%%%%%%%%%%%%%%%%%%%%%%%%%%%%%%%%%%%%%%%%%%%%%%%%%%%%%%%%%%%%%%%%%%

%%%%%%%%%%%%%%%%%%%%%%%%%%%%%%%%%%%%%%%%%%%%%%%%%%%%%%%%%%%%%%%%%%%
\subsection{Domain wall entropy: stripe geometry}
%%%%%%%%%%%%%%%%%%%%%%%%%%%%%%%%%%%%%%%%%%%%%%%%%%%%%%%%%%%%%%%%%%%

To define the domain wall entropy, we consider a $d \times d$ square lattice Ising model in a stripe geometry and introduce two additional spins coupled, respectively, to all of the original spins at the two opposite ends of the stripe. The coupling $\beta$ between these new `boundary' spins $\sigma_L$ and $\sigma_R$ and the bulk spins $\sigma_j$ is the same as between neighboring bulk spins. The conditional correlations $C_{\bm{s}} \equiv \langle \sigma_L \sigma_R\rangle_{\bm{s}}$ between $\sigma_L$ and $\sigma_R$ can be graphically represented by 
\begin{equation}
     \includegraphics[width=.8\columnwidth]{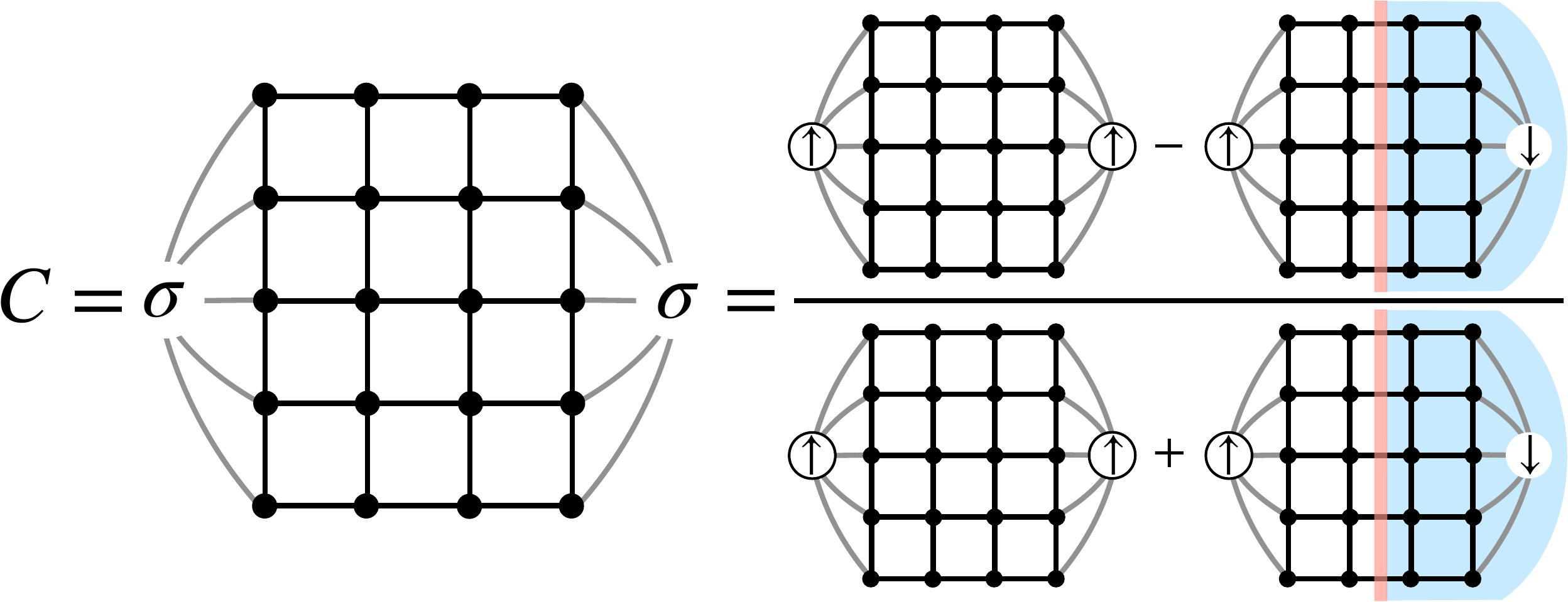} \ .
     \label{eq:domainwall}
\end{equation}

%%%%%%%%%%%%%%%%%%%%%%%%%%%%%%%%%%%%%%%%%%%%%%%%%%%%%%%%%%%%%%%%%%%
\section{Supplementary numerical data}
%%%%%%%%%%%%%%%%%%%%%%%%%%%%%%%%%%%%%%%%%%%%%%%%%%%%%%%%%%%%%%%%%%%

\subsubsection{Moments of domain wall correlation}

\begin{figure}[b!]
    \centering
    \begin{tikzpicture}
        \node[inner sep=0pt] (img) at (0,0) {
            \includegraphics[width=\columnwidth]{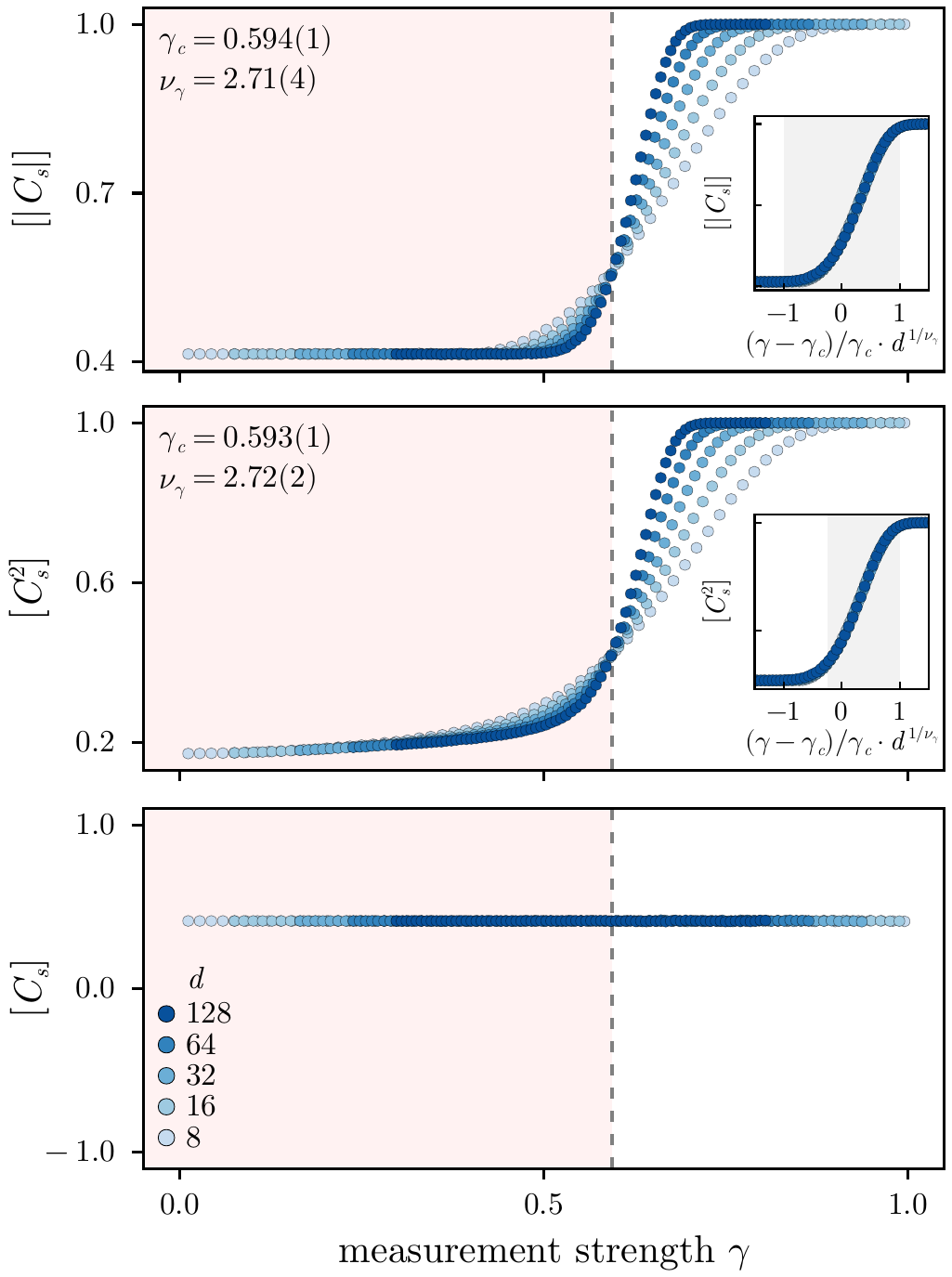}
        };
        \node[fill=white, font=\normalsize, rotate=90] at ([xshift=0.3cm, yshift=9.885cm]img.south west) {$[ |C_{\bm{s}}|]$};
        \node[fill=white, font=\normalsize, rotate=90] at ([xshift=0.3cm, yshift=6.262cm]img.south west) {$[ C_{\bm{s}}^2]$};
        \node[fill=white, font=\normalsize, rotate=90] at ([xshift=0.3cm, yshift=2.625cm]img.south west) {$[ C_{\bm{s}}]$};
        \node[fill=white, font=\scriptsize, rotate=90] at ([xshift=6.535cm, yshift=9.76cm]img.south west) {$[ |C_{\bm{s}}|]$};
        \node[fill=white, font=\scriptsize, rotate=90] at ([xshift=6.535cm, yshift=6.138cm]img.south west) {$[ C_{\bm{s}}^2]$};

    \end{tikzpicture}
    \caption{
        {\bf Varying moments of domain wall correlation functions}, averaged according to the Bayes / Born rule, along the critical line $\beta=\beta_c$ of sweeping $\gamma$.
    }
    \label{fig:criticalcorr}
\end{figure}

Let us supplement the numerical data in the main text by different moments of the domain wall correlation functions $[C_{\bm{s}}]$, $[C_{\bm{s}}^2]$, and $[|C_{\bm{s}}|]$ weeping $\gamma$ along the same critical line $\beta=\beta_c$, see Fig.~\ref{fig:criticalcorr}. The ferromagnetic correlation as a linear average of the density matrix does not depend on the measurement strength $\gamma$, as expected. The disorder average of the absolute value is related to the fidelity correlator~\cite{You24weaksym, Wang24strtowksym} of the classical-quantum mixed state~\cite{Wang25selfdual} $\rho = \sum_{\bm{s}} P(\bm{s}) \ketbra{\bm{s}}\otimes\ketbra{\psi(\bm{s})} $:
\begin{equation}
    [ |C_{\bm{s}}|] = \sum_{\bm{s}} P(\bm{s}) \sqrt{C_{\bm{s}}^2}=\mathcal{F}(\rho, Z_i Z_j \rho Z_j Z_i)\ ,
\end{equation}
where $\mathcal{F}(\rho_1, \rho_2)={\rm tr} \sqrt{\sqrt{\rho_1}\rho_2\sqrt{\rho_1}}$ is the mixed state fidelity. 
$[C_{\bm{s}}^2]$ is the Edwards-Anderson order parameter often employed to detect the spin glass order. The two non-linear probes here signal the phase transition at $\gamma_c$ that are roughly consistent with the entropy.

\begin{figure}[tb]
    \includegraphics[width=\columnwidth]{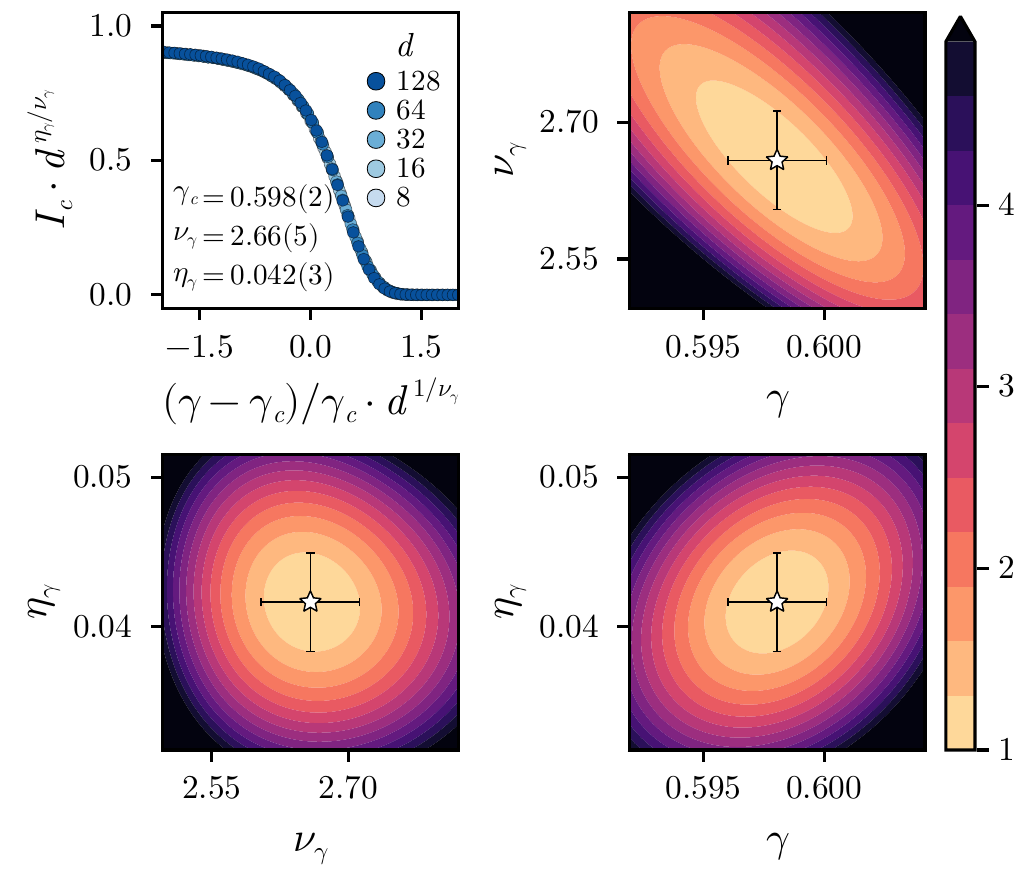}
    \caption{
        {\bf Finite-size data collapse} of the average domain wall entropy/coherent information $I_c$ on the critical line $\beta=\beta_c$
        Besides the critical measurement strength $\gamma_c$ the collapse allows to extract the two critical exponents $\nu_{\gamma}$ and $\eta_{\gamma}$,
        with the three contour plots indicating the quality of the fits in relative coordinates.
    }
    \label{fig:contourplot}
\end{figure}

\subsubsection{Data collapse and critical exponents}

\begin{figure}[htb!]
    \includegraphics[width=\columnwidth]{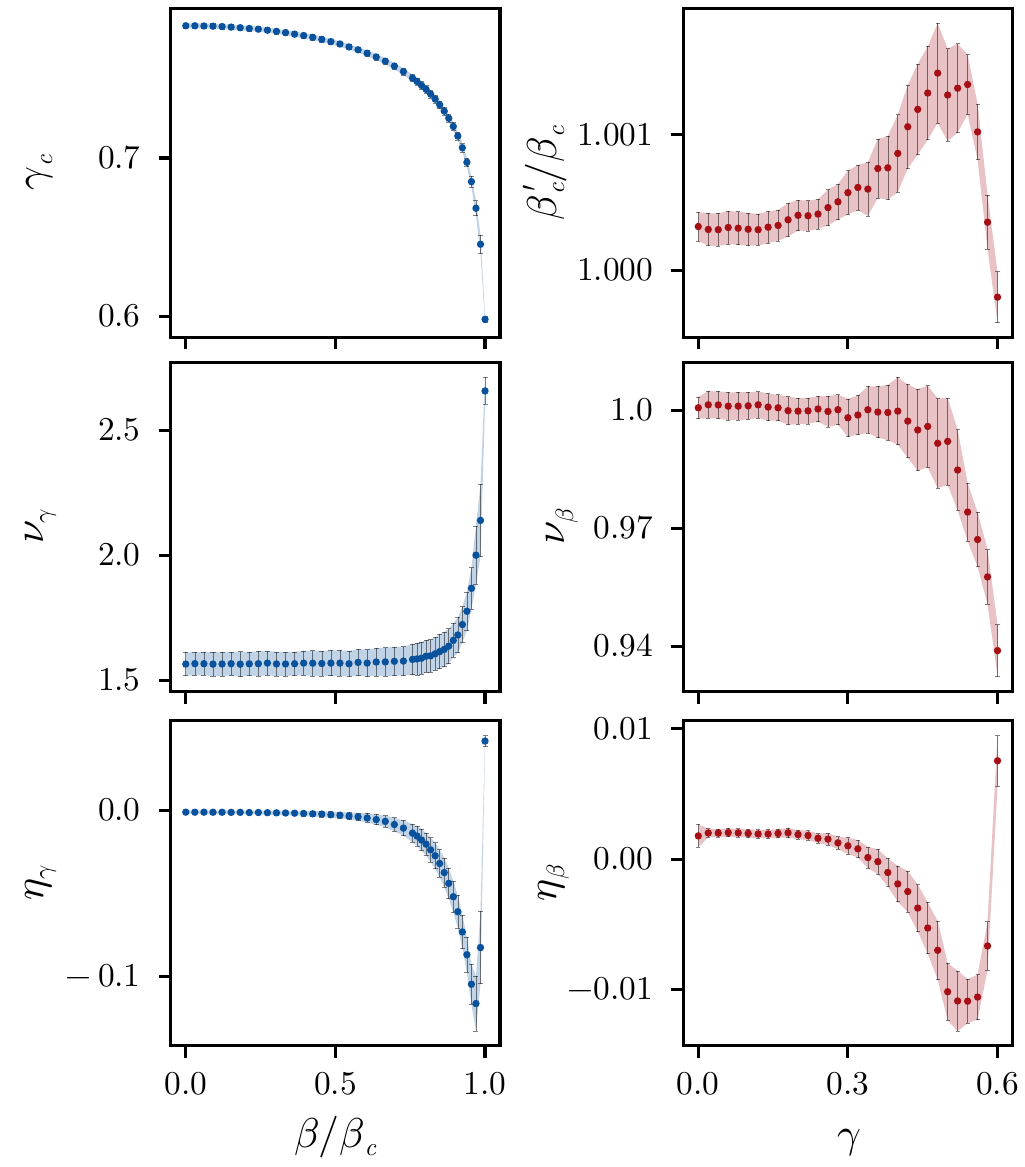}
    \caption{
        {\bf Fit critical exponents along the critical lines}. 
        Left panel: scaling exponents of the inference transition (horizontal cuts of the phase diagram in the main text), for the critical line between the Nishimori criticality at infinite temperature $\beta=0$ and the tricritical point at $\beta_c$.  
        Right panel: scaling exponents of the thermal phase transition (vertical cuts of the phase diagram in the main text), for the critical line between the tricritical point at $\gamma_c$ and the clean Ising point at $\gamma=0$.
        $\beta_c'$ not being exactly $\beta_c$ is a numerical artifact from the finite size scaling, similar to $\eta$ not being exactly $0$ at the Nishimori point (left plot $\beta = 0$) and the clean Ising point (right plot $\gamma = 0$).
    }
    \label{fig:FSS_parameters}
\end{figure}

Next we show a detailed exposition of the finite-size scaling data collapse that we employ to extract the
critical exponents from $I_c$:
\begin{equation}
    I_c(\gamma) = L^{\eta_{\beta}/\nu_{\gamma}} f( d^{1/\nu_{\gamma}}(\gamma-\gamma_c)) \ ,
\end{equation}
where $f(x)$ is a scaling function. 
Fig.~\ref{fig:contourplot} shows such s data collapse for the learning transition at the critical temperature $\beta_c$, when sweeping the measurement strength $\gamma$. 

%%%%%%%%%%%%%%%%%%%%%%%%%%%%%%%%%%%%%%%%%%%%%%%%%%%%%%%%%%%%%%%%%%%

\subsubsection{Critical exponents along the critical lines}

Let us supplement Fig.~\ref{fig:nu} of the main text with detailed fit results for both critical exponents and the critical threshold value 
from the data collapse along the critical lines are shown in Fig.~\ref{fig:FSS_parameters}.

%%%%%%%%%%%%%%%%%%%%%%%%%%%%%%%%%%%%%%%%%%%%%%%%%%%%%%%%%%%%%%%%%%%

\end{document}